\documentclass[a4paper]{article}
\usepackage{graphicx}
\usepackage{amsmath}

\begin{document}

\title{Quantum capacity of channel with thermal noise}
\author{Xiao-yu Chen \\
Lab. of quantum information, China Institute of Metrology, Hangzhou, 310018,
China}
\date{}
\maketitle

\begin{abstract}
The quantum capacity of thermal noise channel is studied. The extremal input
state is obtained at the postulation that the coherent information is convex
or concave at its vicinity. When the input energy tends to infinitive, it is
verified by perturbation theory that the coherent information reaches its
maximum at the product of identical thermal state input. The quantum
capacity is obtained for lower noise channel and it is equal the one shot
capacity.
\end{abstract}

One of the most important issues of classical information theory is the
Shannon formula, which is the capacity of an additive white Gaussian noise
channel. It is achieved when the input is a Gaussian noise source with power
constraint \cite{Shannon} \cite{Cover}.
\begin{equation}
C=\log _2(1+\frac SN),
\end{equation}
where $S$ is the power of the source and $N$ is the power of the noise, the
bandwidth $W$ should be multiplied when it is considered. The formula has
guided the design of the practical communication system for decades.
Correspondingly, in quantum information theory, although a lot of works have
been done \cite{Horodecki}, such a formula is remained to be discovered. The
Shannon formula comes from the Shannon noisy coding theorem, the later gives
the capacity of any noisy channel:
\begin{equation}
C=\sup_XI(X;Y),
\end{equation}
where the supremum is taken over all inputs $X$, $I(X;Y)$ is the Shannon
mutual information and $Y$ is the output. The counterpart of mutual
information in quantum information theory is the coherent information (CI) $%
I_c(\rho ,\mathcal{E})=S(\mathcal{E}(\rho ))-S(\rho ^{RQ^{\prime }})$ \cite
{Schumacher} \cite{Lloyd}. Here $S(\varrho )=-$Tr$\varrho \log _2\varrho $
is the von Neumann entropy, $\rho $ is the input state, the application of
the channel $\mathcal{E}$ resulting the output state $\mathcal{E}(\rho )$; $%
\rho ^{RQ^{\prime }}=$ $(\mathcal{E}\otimes \mathbf{I})(\left| \psi
\right\rangle \left\langle \psi \right| )$, $\left| \psi \right\rangle $ is
the purification of the input state $\rho $. The quantum channel capacity is
\begin{equation}
Q=\lim_{n\rightarrow \infty }\sup_{\rho _n}\frac 1nI_c(\rho _n,\mathcal{E}%
^{\otimes n}).
\end{equation}
The righthand side of above formula was firstly proved to be the upper bound
of quantum channel capacity \cite{Barnum}. The equality was proved at the
postulation of hashing inequality \cite{Horodecki}; the hashing inequality
was lately established \cite{Devetak}. The quantum capacity of a noisy
quantum channel is the maximum rate at which coherent information can be
transmitted through the channel and recovered with arbitrarily good
fidelity. For quantum information channel can be supplemented by one- or
two-way classical channel, thus quantum capacities should be defined with
these supplementary resources. We here deal with the quantum capacity
without any supplementary classical channel\cite{Horodecki}.

Quantum capacity exhibits a kind of nonadditivity \cite{DiVincenzo} that
makes it extremely hard to deal with. Until now, quantum capacity has not
been carried out except for quantum erasure channel\cite{Bennett}. We in
this paper will deal with the quantum capacity of thermal noise quantum
channel (which is addressed as Gaussian quantum channel before \cite
{Harrington}). The general description of the channel is to map the state $%
\rho $ to another state $\mathcal{E}(\rho )$, where $\mathcal{E}$ is a trace
preserving completely positive map. The map $\mathcal{E}$ has a Krauss
operator sum representation. That is $\mathcal{E}(\rho )=\sum_iA_i\rho
A_i^{\dagger }$ with $\sum_iA_i^{\dagger }A_i=I.$ For additive quantum
Gaussian channel, it is quite simple to choose $A_\alpha \ $ to be
proportional to the displacement operator \cite{Chen} \cite{Harrington} $%
D\left( \alpha \right) =$ $\exp [\alpha a^{\dagger }-\alpha ^{*}a]$. The
output state will be
\begin{equation}
\mathcal{E}(\rho )=\frac 1{\pi N}\int d^2\alpha \exp (-\left| \alpha \right|
^2/N)D\left( \alpha \right) \rho D^{\dagger }(\alpha ).  \label{wave1}
\end{equation}
for the simplest situation of thermal noise channel, where $N$ is the
average photon number of the output state if the input is the vacuum.

In dealing with the maximization of the CI, there is a useful lemma in
classical information theory which gives necessary and sufficient conditions
for the global maximum of a convex function of probability distributions in
terms of the first partial derivatives. The lemma was extended to quantum
information theory \cite{Holevo} in evaluating the capacities of bosonic
Gaussian channels. Let $F$ be a convex function on the set of density
operators which contains $\rho _{(0)}$ and $\rho ,$ the necessary and
sufficient condition for $F$ achieves maximum on $\rho _{(0)}$ is that the
convex function $F((1-t)\rho _{(0)}+t\rho )$ of the real variable $t$
achieves maximum at $t=0$ for any $\rho .$ That is $\frac d{dt}\left|
_{t=0}F((1-t)\rho _{(0)}+t\rho )\leq 0\right. $ Generally speaking, CI $%
I_c(\rho ,\mathcal{E})$ is not a global convex function of its input state $%
\rho $. Without lose of generality, let us suppose it is convex \cite{Note0}%
at the vicinity of some $\rho _{(0)}$, the necessary and sufficient
condition that $\rho _{(0)}$ is the maximal state will be
\begin{equation}
\frac d{dt}\left| _{t=0}I_c((1-t)\rho _{(0)}+t\rho ,\mathcal{E})\leq
0\right. .
\end{equation}
where $\rho $ is at the vicinity of $\rho _{(0)}.$ The derivative will be$%
-Tr(\mathcal{E}(\rho )\mathcal{-E}(\rho _{(0)}))\log \mathcal{E}(\rho
_{(0)})+Tr(\rho ^{RQ^{\prime }}-\rho _{(0)}^{RQ^{\prime }})\log \rho
_{(0)}^{RQ^{\prime }}.$ If $\mathcal{E}$ is a trace preserving completely
positive Gaussian operation, then for a gaussian input state $\rho _{(0)}$,
the output state $\mathcal{E}(\rho _{(0)})$ and the joint state $\rho
_{(0)}^{RQ^{\prime }}$ will be Gaussian. Hence their logarithms are
quadratic polynomials in the corresponding canonical variables\cite{Holevo}.
The derivative will be zero under the constraints of the first and second
moments. Where the trace preserving property of $\mathcal{E}$ is also used.
The conclusion is that for input states with the same first and second
moments, Gaussian input state achieves the maximum of CI for a given trace
preserving completely positive Gaussian channel. The same conclusion can be
applied to CI $I_c(\rho _n,\mathcal{E}^{\otimes n}),$ the maximum will be
arrived when the input state is Gaussian.

We then turn to Gaussian input state. Every operators $A\in \mathcal{B}%
\mathcal{(H)}$ is completely determined by its quantum characteristic
function $\chi _A(z):=Tr[AW(z)]$, where $W(z)=\exp [-iz^TR]$ are Weyl
operators and $R=(X_1,P_1,X_2,\cdots ,P_n),$ with $[X_k,P_l]=i\delta _{kl}$.
The density operator $\rho _n$ is called Gaussian, if its characteristic
function $\chi _\rho (z)$ has the form $\chi _\rho (z)=\exp (i\eta ^Tz-\frac
14z^T\gamma z).$ One can show that the first moment $\eta =$Tr$\rho _nR,$the
second moment $\gamma =2$Tr$(R-\eta )\rho _n(R-\eta )^T+iJ_n,$ where
\begin{equation}
J_n=\bigoplus_{k=1}^nJ,\text{ }J=\left[
\begin{array}{ll}
0 & -1 \\
1 & 0
\end{array}
\right]
\end{equation}
The $2n\times 2n$ real symmetric matrix $\gamma $ is usually called the
correlation matrix (CM) of the state $\rho _n$. The first moment represents
the displacement of the state, it is irrelative to the problem of
entanglement as well as channel capacity so that dropped. The completely
positive map on the input state $\rho _n$ will be $\rho _n^{RQ^{\prime }}=(%
\mathcal{E}\otimes \mathbf{I)(}\left| \psi _n\right\rangle \left\langle \psi
_n\right| \mathbf{).}$ The CM of the Schmidt purification $\left| \psi
_n\right\rangle $ is \cite{Holevo}
\begin{equation}
\gamma _\psi =\left[
\begin{array}{ll}
\gamma & \beta \\
\beta ^T & \gamma
\end{array}
\right] ,
\end{equation}
where $\beta =-\beta ^T=J_n\sqrt{-(J_n^{-1}\gamma )^2-\mathbf{I}_{2n}}$ are
purely off-diagonal. It should be noticed that the symplectic eigenvalues of
$\gamma _\psi $ are the square root of the eigenvalues of $-(J_{2\times
n}^{-1}\gamma _\psi )^2,$ where $J_{2\times n}=J_n\oplus (-J_n)$ are chosen
to produce the off-diagonal $\beta $. The application of trace preserving
Gaussian channel will result the state $\rho _n^{RQ^{\prime }}$ with CM \cite
{Giedke}
\begin{equation}
\gamma _\psi ^{\prime }=\left[
\begin{array}{ll}
M_n^T\gamma M_n+N_n & M_n^T\beta \\
\beta ^TM_n & \gamma
\end{array}
\right] ,
\end{equation}
where $M_n=M_1^{\oplus n},N_n=N_1^{\oplus n}.$

For thermal noise quantum channel, we have $M_1=\mathbf{I}_2,N_1=2N\mathbf{I}%
_2.$ Then $\gamma _\psi ^{\prime }$ will be
\begin{equation}
\gamma _\psi ^{\prime }=\left[
\begin{array}{ll}
\gamma +2N\mathbf{I}_{2n} & \beta  \\
\beta ^T & \gamma
\end{array}
\right] .
\end{equation}
The energy of the input state $\rho _n$ is $E_n=\sum_i(\overline{n}_i+1/2)=$%
Tr$\rho _n\sum_i(a_i^{\dagger }a_i+1/2)=($Tr$\gamma )/4$, where we set the
unit of the energy such that $\hbar \omega =1.$ Under the energy constraint
Tr$\gamma =4E_n$, which state will achieve the maximum of the CI? We suppose
the state having maximum CI is $\rho _{n(0)}$ with its CM $\gamma _{n(0)}=2%
\overline{E}\mathbf{I}_{2n},$ where $\overline{E}=E_n/n$ is the average
energy of each mode of input state. We need to prove that
\begin{equation}
-\text{Tr}((\rho _n^{\prime })\mathcal{-}(\rho _{n(0)}^{\prime }))\log \rho
_{n(0)}^{\prime }+\text{Tr}(\rho _n^{RQ^{\prime }}-\rho _{n(0)}^{RQ^{\prime
}})\log \rho _{n(0)}^{RQ^{\prime }}\leq 0,
\end{equation}
where $\rho _n^{\prime }=\mathcal{E}^{\otimes n}(\rho _n),$ $\rho
_{n(0)}^{\prime }=\mathcal{E}^{\otimes n}(\rho _{n(0)}).$ The density
operator $\rho _{n(0)}$ now is the direct product of $\rho _{1(0)}$, $\rho
_{n(0)}=\rho _{1(0)}^{\otimes n}$, we have $\rho _{n(0)}^{\prime }=\rho
_{1(0)}^{\prime \otimes n}$ and $\rho _{n(0)}^{RQ^{\prime }}=\rho
_{1(0)}^{RQ^{\prime }\otimes n}$ as well. The CM\ of $\rho _{n(0)}^{\prime }$
now is $\gamma _{n(0)}+2N\mathbf{I}_{2n}=2(\overline{E}+N)\mathbf{I}_{2n}$,
thus $\rho _{n(0)}^{\prime }=\bigotimes_i(1-v^{\prime })v^{\prime
a_i^{\dagger }a_i}$ is a thermal state with $v^{\prime }=(\overline{E}%
+N-1/2)/(\overline{E}+N+1/2).$ We have $-$Tr$((\rho _n^{\prime })\mathcal{-}%
(\rho _{n(0)}^{\prime }))\log \rho _{n(0)}^{\prime }=-$Tr$((\rho _n^{\prime
})\mathcal{-}(\rho _{n(0)}^{\prime }))\sum_ia_i^{\dagger }a_i\log v^{\prime
}=0$ under the energy constraint Tr$\rho _n\sum_i(a_i^{\dagger }a_i+1/2)=E_n,
$ where Tr$\rho _n^{\prime }\sum_i(a_i^{\dagger }a_i+1/2)=\frac 14$Tr$%
(\gamma +2N\mathbf{I}_{2n})$ and Tr$\rho _n^{\prime }=$Tr$\rho
_{n(0)}^{\prime }=1$ are used. The density operator $\rho
_{1(0)}^{RQ^{\prime }}$ could be diagonalized by some unitary transformation
$U_1,$ the corresponding symplectic transformation
\begin{equation}
S_1=\left[
\begin{array}{ll}
\cosh r\mathbf{I}_2 & -\sinh rJ \\
\sinh rJ & \cosh r\mathbf{I}_2
\end{array}
\right]
\end{equation}
will diagonalize the CM of $\rho _{1(0)}^{RQ^{\prime }},$ that is $S_1\gamma
_{1(0)}S_1^T=\widetilde{\gamma }_{1(0)},$ meanwhile $S_1J\oplus
(-J)S_1^T=J\oplus (-J).$ Here $\tanh 2r=$ $\sqrt{4\overline{E}^2-1}/(2%
\overline{E}+N).$ The diagonalized CM $\widetilde{\gamma }%
_{1(0)}=diag\{\gamma _A,\gamma _A,\gamma _B,\gamma _B\}.$ The density
operator $\rho _{n(0)}^{RQ^{\prime }}$ is a direct product of $\rho
_{1(0)}^{RQ^{\prime }}.$ The unitary transformation diagonalizes $\rho
_{n(0)}^{RQ^{\prime }}$ will be $U_n=U_1^{\otimes n}.$ The corresponding
symplectic transformation will be
\begin{equation}
S_n=\left[
\begin{array}{ll}
\cosh r\mathbf{I}_{2n} & -\sinh rJ_n \\
\sinh rJ_n & \cosh r\mathbf{I}_{2n}
\end{array}
\right] .
\end{equation}
Thus $U_n\rho _{n(0)}^{RQ^{\prime }}U_n^{\dagger }=\rho _{AB}^{\otimes n},$
where $\rho _{AB}=(1-v_A)v_A^{a^{\dagger }a}\otimes $ $(1-v_B)v_B^{b^{%
\dagger }b}$ is a thermal state, with $b,b^{\dagger }$ being the
annihilation and creation operators of `reference' R system which is
introduced in the purification, and $v_j=(\gamma _j-1)/(\gamma _j+1),$ $%
(j=A,B).$ Hence Tr$(\rho _n^{RQ^{\prime }}-\rho _{n(0)}^{RQ^{\prime }})\log
\rho _{n(0)}^{RQ^{\prime }}=$Tr$(U_n\rho _n^{RQ^{\prime }}U_n^{\dagger
}-\rho _{AB}^{\otimes n})\log \rho _{AB}^{\otimes n}$ $=$Tr$(U_n\rho
_n^{RQ^{\prime }}U_n^{\dagger }-\rho _{AB}^{\otimes n})\sum_i(a_i^{\dagger
}a_i\log v_A+b_i^{\dagger }b_i\log v_B).$ After the unitary transformation,
the density operator $U_n\rho _n^{RQ^{\prime }}U_n^{\dagger }$ is an
operator function of the creation and annihilation operators $a_i$,$a_i$,$%
b_i^{\dagger }$ and $b_i^{\dagger }.$ The CM of density operator $U_n\rho
_n^{RQ^{\prime }}U_n^{\dagger }$ will be $S_n\gamma _\psi ^{\prime }S_n^T,$
denote it as
\begin{equation}
\gamma _U\equiv \left[
\begin{array}{ll}
\gamma _{AA} & \gamma _{AB} \\
\gamma _{BA} & \gamma _{BB}
\end{array}
\right] ,
\end{equation}
with $\gamma _{AA}=\cosh ^2r(\gamma +2N\mathbf{I}_{2n})-\sinh ^2rJ_n\gamma
J_n+\sinh r\cosh r(\beta J_n-J_n\beta ^T),$ $\gamma _{BB}=\cosh ^2r\gamma
-\sinh ^2rJ_n(\gamma +2N\mathbf{I}_{2n})J_n+\sinh r\cosh r(J_n\beta -\beta
^TJ_n).$ From the definition of CM, one can get that Tr$U_n\rho
_n^{RQ^{\prime }}U_n^{\dagger }\sum_i((a_i^{\dagger }a_i+1/2)\log
v_A+(b_i^{\dagger }b_i+1/2)\log v_B)=$ $($Tr$\gamma _{AA}\log v_A+$Tr$\gamma
_{BB}\log v_B)/4,$ Tr$\gamma _{AA}=$ $\cosh 2r($Tr$\gamma )+4nN\cosh ^2r$ $%
-\sinh 2r$ Tr$\sqrt{-(J_n^{-1}\gamma )^2-\mathbf{I}_{2n}},$ Tr$\gamma
_{BB}=\cosh 2r($Tr$\gamma )+4nN\sinh ^2r-\sinh 2rTr\sqrt{-(J_n^{-1}\gamma
)^2-\mathbf{I}_{2n}}.$ When the energy is constrained, Tr$\gamma =$Tr$\gamma
_{n(0)},$ we have Tr$(\rho _n^{RQ^{\prime }}-\rho _{n(0)}^{RQ^{\prime
}})\log \rho _{n(0)}^{RQ^{\prime }}=-\frac 14\log (v_Av_B)\sinh 2r($Tr$\sqrt{%
-(J_n^{-1}\gamma )^2-\mathbf{I}_{2n}}-$ $2n\sqrt{4\overline{E}^2-\mathbf{1}}%
).$ For $v_j=(\gamma _j-1)/(\gamma _j+1)<1,$thus$-\log (v_Av_B)>0.$ What
left to be proved is that at the constraint of Tr$\gamma =4n\overline{E}=4E_n
$, Tr$\sqrt{-(J_n^{-1}\gamma )^2-\mathbf{I}_{2n}}$ reaches its maximum when $%
\gamma =\gamma _{n(0)}=2\overline{E}\mathbf{I}_{2n}.$ We start with any
given $\gamma $ with Tr$\gamma =4E_n$. $\gamma $ can be symplectically
diagonalized to $S\gamma S^T=diag\{\gamma _1,\gamma _1,\gamma _2,\gamma
_2,\cdots ,\gamma _n,\gamma _n\}.$ The symplectical transformation $S$ can
be written as $R_2DR_1,$where $R_1,R_2$ are rotations and $D=diag\{d_1,1/d_1,
$ $d_2,1/d_2$ $,\cdots ,$ $d_n,1/d_n\}$ is the squeezing operation. The
rotation does not change the trace of the CM, the squeezing operation
reduces the trace of the CM in the diagonalizing process. Hence Tr$\gamma
\geq 2\sum_i\gamma _i.$ Let $\kappa =($Tr$\gamma )/(2\sum_i\gamma _i)\geq 1,$
$\gamma ^{\prime }=\kappa \cdot diag\{\gamma _1,\gamma _1,\gamma _2,\gamma
_2,\cdots ,\gamma _n,\gamma _n\}$ so that Tr$\gamma =$ Tr$\gamma ^{\prime }.$
We have Tr$\sqrt{-(J_n^{-1}\gamma )^2-\mathbf{I}_{2n}}=$ $2\sum_i\sqrt{%
\gamma _i^2-1}\leq $ Tr$\sqrt{-(J_n^{-1}\gamma ^{\prime })^2-\mathbf{I}_{2n}}%
.$ For the diagonal CM $\gamma ^{\prime }$ with the energy constraint Tr$%
\gamma ^{\prime }=4E_n,$ it is easy to elucidate that when all the diagonal
elements are equal, Tr$\sqrt{-(J_n^{-1}\gamma )^2-\mathbf{I}_{2n}}=$ $2\sum_i%
\sqrt{\gamma _i^{\prime 2}-\mathbf{1}}$ reaches its maximal value $2n\sqrt{4%
\overline{E}^2-\mathbf{1}}.$ We have proved that
\begin{equation}
Tr(\rho _n^{RQ^{\prime }}-\rho _{n(0)}^{RQ^{\prime }})\log \rho
_{n(0)}^{RQ^{\prime }}\leq 0,
\end{equation}
together with Tr$((\rho _n^{\prime })\mathcal{-}(\rho _{n(0)}^{\prime
}))\log \rho _{n(0)}^{\prime }=0.$ Hence $\rho _{n(0)}$ is the extremal
state that maximizes CI as far as CI is convex at the vicinity of $\rho
_{n(0)}$. Similarly, if CI\ is concave at the at the vicinity of $\rho
_{n(0)}$,  $\rho _{n(0)}$ is the extremal state that minimizes CI. CI\ is
the difference of two convex function, quantum mutual information and the
entropy of the input state. This will provide the other way of obtaining the
extremal state.

The next part of this paper is to give evidence of CI\ really reaching its
maximal at $\rho _{n(0)}$ if the input energy is strong enough. The
calculation is based on perturbation theory. Suppose the input state $\rho
_n $ have a complex characteristic function $\chi _n(\mu )=$Tr$(\rho _nD(\mu
))=\chi _{n(0)}(\mu )(1+\varepsilon f(\mu ,\mu ^{*})),$ where $\chi
_{n(0)}(\mu )=\exp [-(N_s+\frac 12)\left| \mu \right| ^2]$ is the complex
characteristic function of $\rho _{n(0)},$ with $N_s=\overline{E}-\frac 12$
being the average photon number of the thermal input $\rho _{1(0)}$ and $\mu
=(\mu _1,\mu _2,\cdots ,\mu _n)$. The perturbation item $f(\mu ,\mu ^{*})$
can be expanded with power of $\mu $ and $\mu ^{*}$. The condition that the
first and second moments of $\rho _n$ are equal to that of $\rho _{n(0)}$
leads to all the item of the linear and square power in the expansion of $%
f(\mu ,\mu ^{*})$ being $0.$ The cubic item $\mu ^{3-i}\mu ^{*i}$ as well as
other odd power items will have no contribution in the first order
perturbation. So the first none zero contribution will be the fourth power.
The other even power items can be neglected comparing with the fourth power
as the input energy become strong enough. So we suppose
\begin{equation}
f(\mu ,\mu ^{*})=\sum_{i\geq j}c_{ij}\left| \mu _i\mu _j\right| ^2.
\end{equation}
It should be noted that items such as $\mu _1\mu _2^{*}\left| \mu _i\right|
^2$ and $\mu _1^2\mu _2^{*2}$ will also contribute to the first order
perturbation, but they can also be neglected at the strong input energy as
we will see below. All other items with unequal number of $\mu $ and $\mu
^{*}$ will not contribute to the first order perturbation.

Lets first consider the $\left| \mu _1\right| ^4$ item, the input state now
is a direct product of perturbed first mode and other $n-1$ thermal state
modes. The situation is reduced to deal with the perturbation problem of $%
\chi _1(\mu _1)=\chi _{1(0)}(\mu _1)(1+\varepsilon \left| \mu _1\right| ^4).$
We have $\rho _1=(1+\varepsilon \frac{d^2}{dN_s^2})\rho _{1(0)}=\rho
_{1(0)}+\varepsilon \phi .$ $\rho _{1(0)}=(1-v_s)\sum_{k=0}^\infty
v_s^k\left| k\right\rangle \left\langle k\right| ,$ with $v_s=N_s/(N_s+1).$
The strict eigenvalues of $\rho _1$ are $\lambda _k=\lambda
_{k(0)}+\varepsilon \phi _k,$ with $\lambda _{k(0)}=(1-v_s)v_s^k$ and $\phi
_k=\lambda _{k(0)}[2-4k/N_s+k(k-1)/N_s^2]$. The entropy of $\rho _1$ can be
expanded up to the second derivative as $S(\rho _1)=S(\rho _{1(0)})-\frac
12\varepsilon ^2\sum_k\phi _k^2/\lambda _{k(0)}+o(\varepsilon ^3),$ where
null first and second moments of $\phi $ are used. The calculation of the
entropy of $\rho _1^{\prime }$ is straightforward, it is $S(\rho _1^{\prime
})=S(\rho _{1(0)}^{\prime })-\frac 12\varepsilon ^2[\frac 2{N^{\prime
}(N^{\prime }+1)}]^2+o(\varepsilon ^3)$ , with $N^{\prime }=N_s+N.$ The
purification of $\rho _{1(0)}$ is $\rho _1^{RQ}=\sum_{km}\sqrt{\lambda
_k\lambda _m}\left| kk\right\rangle \left\langle mm\right| ,$ such a
purification is more frequently used in literature but different from our
above purification$.$ The state $\rho _1^{RQ}$ then is expanded in $%
\varepsilon $ to the linear item\cite{Note1}. $\rho _1^{RQ}=\rho
_{1(0)}^{RQ}+\varepsilon \Phi ,$ with $\Phi =\frac 12(\Phi _0+\Phi
_0^{\dagger }),$ $\Phi _0=(1-v_s)^2[1-4a_1^{\dagger }b_1^{\dagger
}(1-v_s)v_s^{-1/2}+a_1^{\dagger 2}b_1^{\dagger 2}(1-v_s)^2v_s^{-1}]\rho
_{1(0)}^{RQ}.$ It can be proved that
\begin{equation}
(\mathcal{E}\otimes \mathbf{I)}a_1^{\dagger j}b_1^{\dagger j}\rho
_{1(0)}^{RQ}=v_s^{-j/2}a_1^{\dagger j}a_1^j\rho _{1(0)}^{RQ^{\prime }}.
\end{equation}
Thus $\rho _1^{RQ^{\prime }}=\rho _{1(0)}^{RQ^{\prime }}+\varepsilon \Phi
^{\prime },$ with $\Phi ^{\prime }=\frac 12(\Phi _0^{\prime }+\Phi
_0^{\prime \dagger }),$ $\Phi _0^{\prime }=(1-v_s)^2[2-4a_1^{\dagger
}a_1/N_s+a_1^{\dagger 2}a_1^2/N_s^2]\rho _{1(0)}^{RQ^{\prime }}.$ Note that
the trace and all first and second moment of $\Phi ^{\prime }$ are null. The
eigenstates of $\rho _{1(0)}^{RQ^{\prime }}$ are $V_1\left| km\right\rangle $
with eigenvalues $\lambda _{km(0)}=(1-v_A)v_A^k$ $(1-v_B)v_B^m$, where $V_1$
diagonalizes $\rho _{1(0)}^{RQ^{\prime }}$ and
\begin{eqnarray}
V_1a_1V_1^{\dagger } &=&a_1\cosh r-b_1^{\dagger }\sinh r, \\
V_1b_1V_1^{\dagger } &=&b_1\cosh r-a_1^{\dagger }\sinh r,  \nonumber
\end{eqnarray}
with $\tanh 2r=2\sqrt{N_s(N_s+1)}/(N_s+N^{\prime }+1).$ The first order
perturbation to the eigenvalue will be $\Phi _{km}^{\prime }=\left\langle
km\right| V_1^{\dagger }\Phi ^{\prime }V_1\left| km\right\rangle $ which is
\begin{eqnarray}
\Phi _{km}^{\prime } &=&\lambda _{km(0)}(1-v_s)^2\{2-4[k\cosh ^2r+(m+1)\sinh
^2r]/N_s \\
&&+[k(k-1)\cosh ^4r+(m+1)(m+2)\sinh ^4r+4k(m+1)\sinh ^2r\cosh ^2r]/N_s^2\}.
\nonumber
\end{eqnarray}
Up to $\varepsilon ^2$ item, the entropy will be $S(\rho _1^{RQ^{\prime
}})\approx S(\rho _{1(0)}^{RQ^{\prime }})-\frac 12\varepsilon
^2\sum_{km}\Phi _{km}^{\prime 2}/\lambda _{km(0)}.$ After the summation and
taking the limitation of $N_s\rightarrow \infty ,$ the total increase of CI
between the input $\rho _n$ and $\rho _{n(0)}$ will be
\begin{equation}
\lim_{N_s\rightarrow \infty }[I_c(\rho _n)-I_c(\rho _{n(0)})]=-\frac
12\varepsilon ^2[\frac 4{N_s^4}-\frac 3{2N_s^4}]<0.
\end{equation}
the positive part which comes from $\rho _n^{RQ^{\prime }}$ state is only $%
\frac 38$ of the negative part which comes from $\rho _n^{^{\prime }}$. Thus
$I_c(\rho _{n(0)})$ is maximal in this situation.

The next perturbation item is $\left| \mu _1\mu _2\right| ^2$. We only need
to deal with the first and second modes with $\chi _2(\mu _1,\mu _2)=\chi
_{2(0)}(\mu _1,\mu _2)(1+\varepsilon \left| \mu _1\mu _2\right| ^2)$ \cite
{Note2}while other modes are kept in thermal states and irrelative. Here the
degenerate perturbation is applied. The calculation of the entropy
difference of $\rho _2^{^{\prime }}$ and $\rho _{2(0)}^{\prime }$ is easy
because the perturbation operator is diagonal in degenerate subspace. The
result is
\begin{equation}
S(\rho _2^{^{\prime }})-S(\rho _{2(0)}^{\prime })=-\frac{\varepsilon ^2}{%
2N^{^{\prime }2}(N^{\prime }+1)^2}.
\end{equation}
The calculation of the entropy difference of $\rho _2^{RQ^{\prime }}$ and $%
\rho _{2(0)}^{RQ\prime }$ will encounter with non diagonal operators $%
a_1a_2^{\dagger }b_1b_2^{+}$ and $a_1^{\dagger }a_2b_1^{\dagger }b_2$ in
degenerate subspace which indicate inter-mode particle transfer, but in the
degenerate subspace the total particle number of Q system (or R system) is
conserved . The entropy difference can be calculated by first summing up in
the degenerate subspace then the total particle number of Q system and R
system. In the summation Tr$M_{k_1m_1}^{l2}$ is involved, where $M^l$ is the
perturbation operator in the $lth$ degenerate subspace, fortunately it is $%
\sum_{k_1m_1}M_{k_1m_1}^{l2}$ by the special structure of $M^l.$ The final
result after the summation and taking the limitation of $N_s\rightarrow
\infty $ will simply be
\begin{equation}
\lim_{N_s\rightarrow \infty }S(\rho _2^{RQ^{\prime }})-S(\rho
_{2(0)}^{RQ^{\prime }})=-\frac{3\varepsilon ^2}{16N_s^4}.
\end{equation}
Still it is $\frac 38$ of the entropy difference of $\rho _2^{^{\prime }}$
and $\rho _{2(0)}^{\prime }$. Thus we have
\begin{equation}
\lim_{N_s\rightarrow \infty }[I_c(\rho _n)-I_c(\rho _{n(0)})]=-\frac{%
5\varepsilon ^2}{16N_s^4}<0.
\end{equation}

In the situation of $\chi _2(\mu _1,\mu _2)=\chi _{2(0)}(\mu _1,\mu
_2)[1+\varepsilon (\left| \mu _1\right| ^2+c\left| \mu _1\mu _2\right| ^2)]$
, the entropy difference will be the sum of each term because the cross item
is null by the null of the first and second moment of $\phi $ and $\Phi
^{\prime }.$ The general case of $f(\mu ,\mu ^{*})$ will be
\begin{equation}
\lim_{N_s\rightarrow \infty }[I_c(\rho _n)-I_c(\rho _{n(0)})]=-\frac{%
5\varepsilon ^2}{16N_s^4}(4\sum_ic_{ii}^2+\sum_{i\neq j}c_{ij}^2)<0.
\end{equation}

The conclusion is that thermal state of infinitive energy achieves the
maximal of coherent information. That is
\begin{equation}
\lim_{n\rightarrow \infty }\max_{\rho _n}\frac 1nI_c(\rho _n,\mathcal{E}%
^{\otimes n})=\max \{0,-\log _2(eN)\},
\end{equation}
( $e=2.71828\ldots $).The result is verified up to the nonzero lowest power
of the inverse of the input energy in each channel use. Whether it is
correct for the state without an item contributing to $N_s^{-4}$ is not
known.

We obtain a local maximum, whether it is the global supremum should be
verified. For lower noise channel, this can be verified. For a given
channel, the CI difference is a function of $N_s.$ In the two case we
calculated, the CI difference is positive infinitive at $N_s\rightarrow 0,$
as $N_s$ increases, it monotonically decreases to $0.$ After that it
decreases further to negative then increases but still keeps negative and
never turns to positive. At $N_s\rightarrow \infty ,$ it is negative as we
elucidated above. Denote the zero point as $N_{s0},$ then calculate the
quantum mutual information $I(\rho _{n(0)}(N_{s0}),N)$ which is the supremum
of all state with equal or less energy. $I(\rho _n,N)$ is greater $I_c(\rho
_n,N).$ Hence if $I(\rho _{1(0)}(N_{s0}),N)\leq -\log _2(eN)$ $,$ we have
CI\ to be less than $-n\log _2(eN)$ in the interval $[0,N_{s0}]$, thus for $%
N\leq N_c$ we can safely conclude that
\begin{equation}
Q=\max \{0,-\log _2(eN)\},
\end{equation}
as far as the high order perturbation and high power of $\mu $ do not
destroy the maximal property of CI at infinitive input energy, where $N_c$
is the solution of $I(\rho _{1(0)}(N_{s0}),N_c)=-\log _2(eN_c)$ . We have $%
N_c=0.1756$ which comes from the two-mode perturbation.

For the $n$ uses of the thermal noise channel with $N\leq N_c$, the supremum
of the whole CI\ is achieve by an input of the direct product of the
identical thermal noise states. It is followed that the quantum capacity of
thermal noise channel is equal to the one-shot quantum capacity \cite{Holevo}%
\cite{Chen} of the channel. The achievable of the quantum capacity of the
thermal noise channel by quantum error-correction codes had been proven\cite
{Harrington}.

\textit{\ }Funding by the National Natural Science Foundation of China
(under Grant No. 10575092), Zhejiang Province Natural Science Foundation
(under Grant No. RC104265) and AQSIQ of China (under Grant No. 2004QK38) are
gratefully acknowledged.

\end{document}